\documentclass[preprint,aps,prb]{revtex4}

\usepackage{bm}
\def\vec#1{\bm#1}

\usepackage{graphicx}

\begin{document}

\title{Charge ordering in the intermediate valence magnet YbPd}
\author{Ryo~Takahashi$^1$, Takashi~Honda$^1$, Atsushi~Miyake$^2$, Tomoko~Kagayama$^2$, Katsuya~Shimizu$^2$, Takao~Ebihara$^3$, Tsuyoshi~Kimura$^1$, and Yusuke~Wakabayashi$^1$\footnote{e-mail: wakabayashi@mp.es.osaka-u.ac.jp}}
\address{$^1$Division of Materials Physics, Graduate School of Engineering Science,\\
 Osaka University, Toyonaka 560-8531, Japan\\
$^2$ Center for Quantum Science and Technology under Extreme Conditions, Osaka University, Osaka 560-8531, Japan\\
$^3$ Department of Physics, Faculty of Science, Shizuoka University, Ohya, Shizuoka 422-8529, Japan 
}

\date{\today}
\begin{abstract}
An x-ray diffraction study reveals the charge ordering structure in an intermediate valence magnet YbPd with a CsCl-structure. The valence of the Yb ions forms an incommensurate structure, characterized by the wavevector ($\pm$0.07 $\pm$0.07 1/2) below 130~K. At 105~K, the incommensurate structure turns into a commensurate structure, characterized by the wavevector (0 0 1/2). Based on the resonant x-ray diffraction spectra of the superlattice reflections, the valences of the Yb ions below 105~K are found to be 3+ and 2.6+. 
The origin of the long wavelength modulation is discussed with the aid of an Ising model having the second nearest neighbor interaction.
\end{abstract}
\maketitle

\section{Introduction}

Electronic charges sometimes show intriguing behavior induced by their mutual interactions in condensed matter. Charge ordering is one of the most widely studied phenomena in transition metal oxides\cite{MItransition}, organic conductors\cite{org} and rare earth compounds\cite{Ochiai90JPSJ,Staub05PRB,Inami10PRB}. Normally, charge ordering is seen in materials with low conductivity, and the phenomenon involves a sudden increase in the electrical resistivity. However, YbPd is a good metal in all temperature ranges\cite{Pott85PRL} with resistivity of $\sim 10^2~\mu \Omega$cm, and is proposed to have a charge order accompanied by a decrease in resistivity.\cite{Bonville86PRL} 

YbPd is an interesting compound in many ways. 
In the Yb-Pd phase diagram, this CsCl-structured intermetallic compound stands at the border of Yb$^{2+}$ and Yb$^{3+}$, which are stabilized in the Yb rich side and the Pd rich side, respectively\cite{Iandelli73RevChimMiner}.
The valence of the Yb ions in YbPd measured by $L_{III}$-edge x-ray absorption spectroscopy was actually fractional 2.8+ across all temperature ranges\cite{Pott85PRL}. In the fractional valence state, it shows magnetic ordering at $T_{\rm N}=$1.9~K.\cite{Bonville86PRL}
Magnetic ordering in the fractional valence state is not very common, and we were therefore interested in studying the origin of this rare phenomenon. 
 M\"ossbauer measurements below 4.2~K show two kinds of Yb, one magnetic and the other non-magnetic, coexisting in equal proportions.\cite{Bonville86PRL} 
There are five phase transitions, at $T_1$=125~K, $T_2$=105~K, $T_{\rm N}=$1.9~K, $T_{\rm MH}=$0.6~K and $T_{\rm ML}=$0.3~K.\cite{Miyake12JPhysConf} While the low temperature transitions below 2~K are known to be magnetic transitions, the other two transitions have not yet been explained. The high temperature transitions are accompanied by strong anomalies in the material's specific heat, thermal expansion\cite{Pott85PRL} and elastic constants\cite{Nakanishi11ChineseJPhys}. Although these observations collectively suggest that either $T_1$ or $T_2$ is the charge ordering temperature, no superstructure has been reported to date.

Charge ordering can be studied by ordinary x-ray diffraction because of the difference in the ionic radius of Yb$^{2+}$ and Yb$^{3+}$, which produces a Pd displacement in the charge ordering phase. A more relevant probe for the spatial valence arrangement is resonant x-ray diffraction, which is a combination of an x-ray diffraction technique with spectroscopy. Using the $L_{III}$ absorption edge energy, it is possible to study the valence arrangements of $4f$ electron systems\cite{Staub05PRB,Inami10PRB}. We have succeeded in clearly observing the superstructure induced by charge ordering in YbPd by means of these bulk sensitive techniques.

\section{Experiment}
Single crystal samples were grown by the self-flux method.\cite{Canfield92PhilMag} The as-grown samples have facets parallel to the \{001\}-planes and are cube-shaped. The typical size of these crystals is 1~mm$^3$. 
 Non-resonant x-ray diffraction measurements were performed with a four-circle x-ray diffractometer attached to a Mo $K_\alpha$ x-ray generator. The incident x-ray beam was monochromatized with a bent graphite monochromator, and the scattered beam was then measured with a charge-coupled device (CCD) camera or a point detector. 
 To find the superlattice reflections, x-ray photographs were taken with 18~keV intense synchrotron radiation at the BL-8B beamline of the Photon Factory, KEK, Japan. 
The charge order was examined by the resonant x-ray scattering technique at the Yb $L_{III}$ absorption edge. The measurement was performed at the BL-4C beamline of the Photon Factory with a four-circle diffractometer. The sample temperature was controlled by using a closed cycle refrigerator for all measurements. 

\section{Results and analysis}
\subsection{Lattice parameters}
\begin{figure}
\includegraphics[width=8cm]{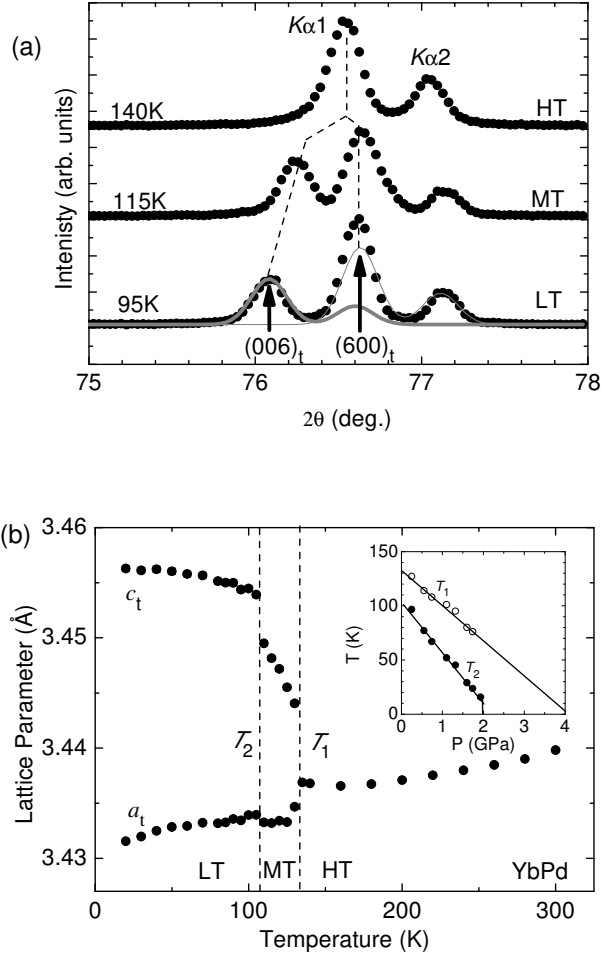}
\caption{(a)$\theta$-2$\theta$ line profile of (006)$_{\rm c}$ Bragg reflection at 140~K, 115~K and 95~K measured with Mo $K_\alpha$ x-rays. The error bars are shorter than the symbol size. The thick and thin gray curves for the 95~K profile show the results of the peak separation using a double Gaussian for the $K_{\alpha 1,2}$ doublet. The dashed lines are intended as visual guides.
(b) Temperature variation of the lattice parameters. Inset: Phase diagram reported in ref.{\protect{\onlinecite{Miyake12JPhysConf}}}.}
\label{fig:006}
\end{figure}

Figure~\ref{fig:006}(a) shows the $\theta$-2$\theta$ line profiles of (006)$_{\rm c}$ Bragg reflection at 140~K (above $T_1$, high temperature (HT) phase), 115~K (between $T_1$ and $T_2$, medium temperature (MT) phase) and 95~K (below $T_2$, low temperature (LT) phase) measured with the Mo $K_\alpha$ x-rays. Here, the suffix c denotes the index in the high-temperature cubic phase. In addition to the doublet caused by the $K_{\alpha 1}$ and $K_{\alpha 2}$ x-rays, clear peak splitting with an intensity ratio of 1:2 was found below $T_1$. This result indicates that the highly symmetrical YbPd deforms into a tetragonal structure below $T_1$, and the single crystal turns into a multi-domain crystal.
Panel (b) shows the lattice parameters $a_{\rm t}$ and $c_{\rm t}$ obtained from the peak positions of the (600)$_{\rm t}$ and (006)$_{\rm t}$ reflections, where the suffix t denotes the index in the tetragonal phase. Based on the peak profile, we found that $T_1$ is between 130~K and 135~K. Above $T_1$, $a_{\rm t}$ is equal to $c_{\rm t}$ because the HT phase is cubic. Both of these transitions induce volume expansions with cooling. This behavior agrees well with the slope of the phase boundary in the pressure-temperature phase diagram shown in the inset of panel (b).\cite{Miyake12JPhysConf}  

The isotropic atomic displacement parameters given by $B=8\pi\langle u^2 \rangle$, where $u$ denotes the atomic displacement from the equilibrium position, for Yb and Pd at 300~K were estimated from the $(00l)_{\rm c}$ intensities. The parameter values were 0.7~\AA$^2$ and 1.4~\AA$^2$, respectively. Because the Lindemann melting criterion predicts a value of 0.8~\AA$^2$ for $B$ at 300~K, the $B$ for Pd is exceptionally large. The large $B$ value for Pd may thus originate from the fluctuation of the Yb radius caused by the valence fluctuation.

\subsection{Crystal structure and charge ordering in the LT phase}
\begin{figure}
\includegraphics[width=8cm]{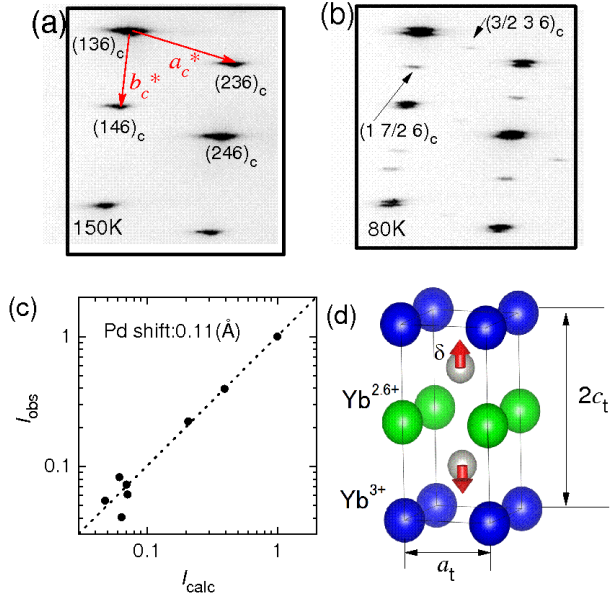}
\caption{
(Color online) Oscillation photographs around (136)$_{\rm c}$ taken in (a) the HT phase (150~K) and (b) the LT phase (80~K). (c) Observed intensity at 95~K plotted against the calculated intensity for the best fitting structure. (d) Schematic view of the LT phase structure.}
\label{fig:IP}
\end{figure}

The synchrotron diffraction experiment was performed to search for superlattice reflections. Figure~\ref{fig:IP} shows the oscillation photographs around (136)$_{\rm c}$ taken in (a) the HT phase, and (b) the LT phase. Along with the peak splitting caused by the cubic-tetragonal phase transition, the superlattice reflections characterized by the wavevector ($00\frac12$)$_{\rm c}$ were observed. 

The positions of the observed superlattice reflections were examined by using the four-circle diffractometer. The superlattice reflections in the LT phase were found to be at $(00\frac12)_{\rm t}$ away from the fundamental Bragg reflections and no reflections were seen at the $(\frac1200)_{\rm t}$ positions. The peak widths of the superlattice reflections were the same as those of the fundamental Bragg reflections, which means that the correlation of the superstructure reaches a long range. A strong (0 0 6.5)$_{\rm t}$ reflection was observed, whereas (6 0 $\frac12$)$_{\rm t}$ reflection was very weak. This feature indicates that the atomic displacement $\vec \delta$ in the LT phase is parallel to the $c$-axis, because the superlattice reflection intensity caused by the small atomic displacement is proportional to $|\vec Q \cdot \vec \delta|^2$, where $\vec Q$ denotes the scattering vector.\cite{Qd} We measured the Bragg and superlattice intensities along the (00$l$)$_{\rm t}$ line at 95~K to obtain the structure in the LT phase. Figure~\ref{fig:IP}(c) shows the integrated intensities of the (00$l$)$_{\rm t}$ peaks plotted against those fitted by the $B$s and the $\vec \delta$s along the $c$-axis for both Pd and Yb. 
It was found that the Pd atoms are displaced by 0.11\AA\/ along the $c$-direction to form a twofold structure in the LT phase, as shown in Fig.~\ref{fig:IP} (d). No atomic displacement was found for the Yb ions. This structure implies an alternate arrangement of high- and low-valence Yb ions along the $c$-direction.

To clarify the valence arrangement, the energy spectra of the superlattice reflections were measured. Each element has a characteristic absorption edge, and the atomic form factor $f$ strongly depends on the x-ray energy $E$ around the edge. Therefore, $f$ is written as $f(Q,E)=f_0(Q)+f'(E)+if''(E)$, where $f_0$ denotes the Thomson scattering term, and $f'$ and $f''$ denote the real part and the imaginary part of the anomalous dispersion term, respectively. 
The difference in the edge energies between the divalent and trivalent Yb ions is reported to be 7~eV.\cite{Staub05PRB} This energy difference enables us to distinguish the valence values of these ions.

\begin{figure}
\includegraphics[width=8cm]{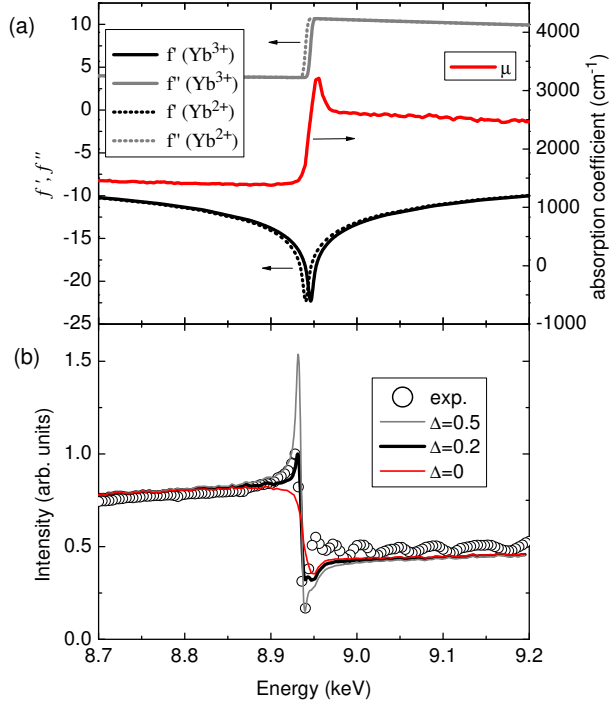}
\caption{
(Color online) (a) Anomalous dispersion terms $f'$ and $f''$ for Yb$^{2+}$ and Yb$^{3+}$ based on Hartree-Fock calculations.\protect{\cite{Sasaki_table}} The linear absorption coefficient that was estimated from the fluorescence spectrum is also shown.
(b) Energy spectrum of the ($00\frac 12$) superlattice reflection measured at 80~K. The gray, thick solid, and thin solid curves show the calculated spectra for $\Delta$=0.5, 0.2 and 0, respectively. }
\label{fig:RXS}
\end{figure}

 The structure factor $F$ for the superlattice reflections ($00\frac L2$)$_{\rm t}$ is written as
\begin{eqnarray}
F_{00\frac L2}&=&
(-1)^{(L-1)/2} 2f_{\rm Pd} \sin (4 \pi L \delta) + f_{H} - f_{L}\label{eq:F},\\
f_H&=& (n_f+\Delta) f_{3+} + \{1-(n_f+\Delta)\} f_{2+},\\
f_L&=& (n_f-\Delta) f_{3+} + \{1-(n_f-\Delta)\} f_{2+},
\end{eqnarray}
where $f_{\rm Pd}$, $f_{3+}$ and $f_{2+}$ denote the atomic form factors for the Pd, Yb$^{3+}$ and Yb$^{2+}$ ions, $\delta$ shows the Pd displacement in the unit of $c_{\rm t}$, $n_f$ is the average hole concentration of the Yb $f$-orbital with a value close to 0.8,\cite{Pott85PRL} and $\Delta$ denotes the amplitude of the charge ordering. The energy dependence of the scattered intensity is $|F|^2 / \mu (E)$, where $\mu(E)$ denotes the absorption coefficient.\cite{Dumesnil98PRB} Using the theoretically calculated anomalous dispersion term for an isolated Yb atom\cite{Sasaki_table} ($f_c(E)$) and the structure factor from eq.(\ref{eq:F}), we analyzed our experimental results. The values of $f_{3+}$ and $f_{2+}$ are obtained from $f_0(Q)+f_c(E\pm 3.5 {\rm eV})$, and $\mu (E)$ was extracted from the fluorescence of the sample (see Fig.\ref{fig:RXS}(a)). The theoretically calculated spectra were convoluted with the experimental resolution function, which was a Gaussian with a full-width at half maximum of 7~eV. The measured energy spectrum for the ($00\frac 12$)$_{\rm t}$ superlattice reflection is shown in Fig. \ref{fig:RXS}(b). The upturn in the intensity with increasing energy toward the absorption edge is not expected without charge ordering, as indicated by the thin solid curve, and is therefore a signature of the charge order.
Based on a comparison between the experimentally observed spectra and the calculated spectra, we conclude that $\Delta \sim 0.2$, i.e., that the high- and low-valence Yb ions are 3+ and 2.6+, respectively, for the LT phase of YbPd, as shown in Fig.~\ref{fig:IP} (d).

\subsection{Incommensurate structure in the MT phase}

\begin{figure}
\includegraphics[width=8cm]{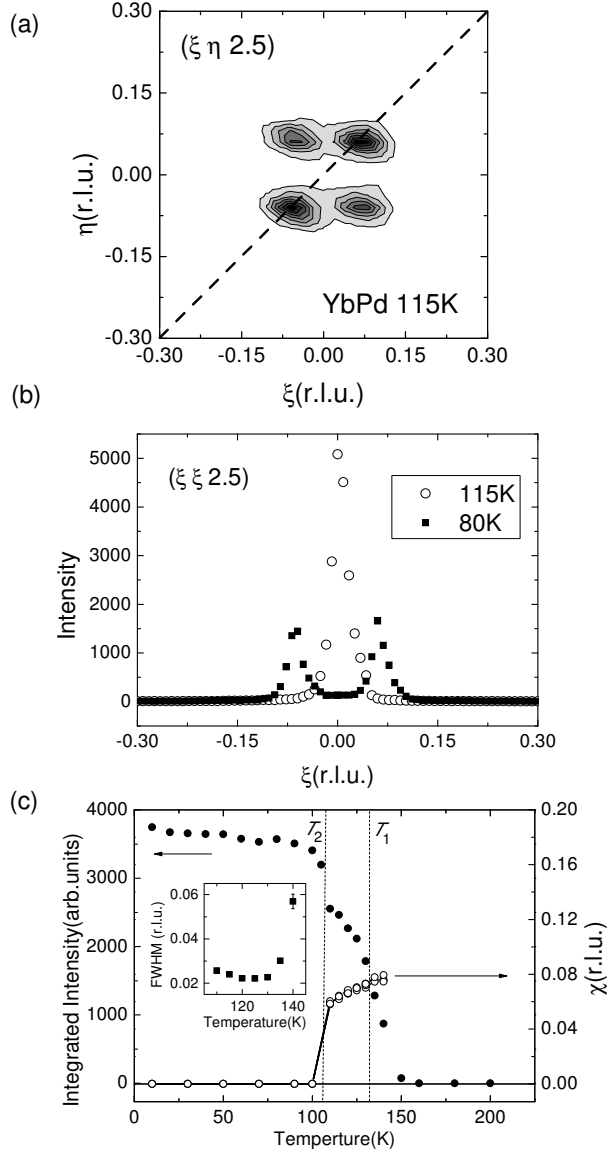}
\caption{
(a) Scattering intensity distribution of the ($\xi \eta 2.5$)$_{\rm t}$ plane at 115~K. (b) Line profiles along the dashed line in panel (a) at 115~K and 80~K.
 (c) Temperature dependence of the incommensurability $x$ and the integrated intensity.
}
\label{fig:map}
\end{figure}

\begin{figure}
\includegraphics[width=16cm]{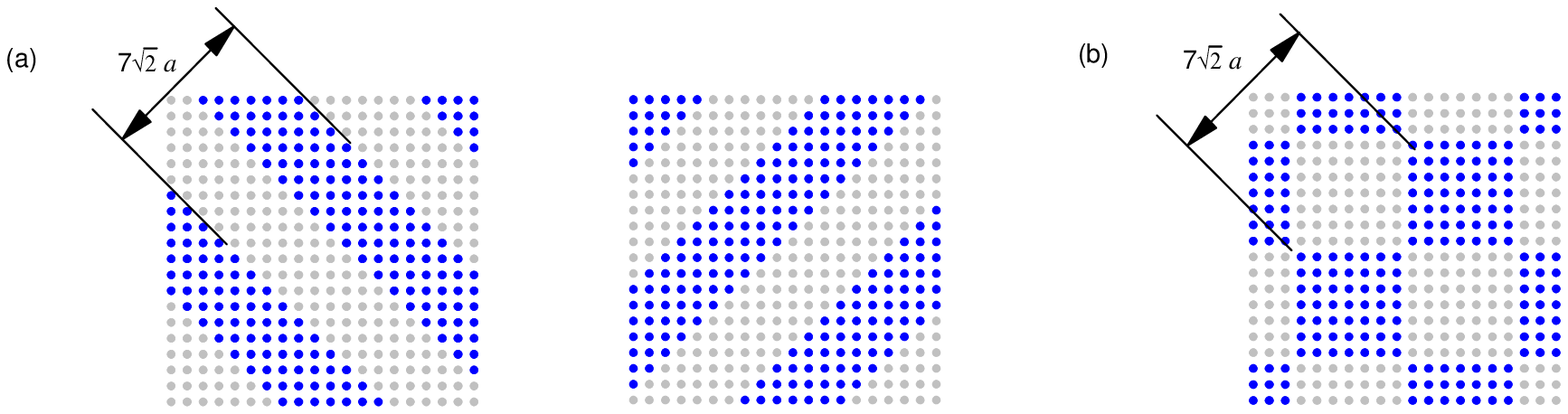}
\caption{(Color online) Schematics of the possible modulation structures in the MT phase. (a) Stripe structure; (b) checkerboard structure. The dark and gray symbols represent the Yb ions with the different valence states.
}
\label{fig:models}
\end{figure}

In the MT phase, we discovered a splitting of the superlattice reflections. Figure~\ref{fig:map}(a) shows the scattering intensity distribution of the ($\xi \eta 2.5$)$_{\rm t}$ plane at 115~K. There are four peaks at ($\pm x$ $\pm x$ 2.5)$_{\rm t}$. Panel (b) shows the line profiles along ($\xi$ $\xi$ 2.5)$_{\rm t}$ (the dashed line in panel (a)) at 115~K (MT phase), and at 80~K (LT phase). The peak position, i.e., the incommensurability $x$, changes from 0.07 in the MT phase to 0 with the phase transition at $T_2$. 
This result indicates that the charge order in the MT phase is an alternating arrangement along the $c$-direction, which is similar to that of the LT phase, but it has an in-plane modulation. Similar to the LT phase, the superlattice peaks are as narrow as those of the Bragg reflections, and therefore a long-range correlation was established. There are two possible modulation structures, depicted by the stripe and checkerboard structures that are presented in Fig.\ref{fig:models}. The stripe structure is characterized by valence modulation along the [$110$]$_{\rm t}$ direction with a wavelength of $\sim 7\sqrt 2 a$. This modulation structure provides two superlattice peaks at ($+x$ $+x$ 2.5)$_{\rm t}$ and ($-x$ $-x$ 2.5)$_{\rm t}$ in the region shown in Fig.~\ref{fig:map} (a). The other two peaks are produced by the twin variant with a modulation vector along [$1\bar 10$]$_{\rm t}$. When we adopt the checkerboard structure, an alternating arrangement of $7a \times 7a$-sized tiles is expected to produce the four peaks shown in Fig.~\ref{fig:map} (a).

The temperature variations of the incommensurability and the intensity integrated over all the peaks around (0 0 2.5)$_{\rm t}$ are presented in panel (c). The MT-phase superlattice intensity was found below 140~K, which is slightly above $T_1$. The inset shows that the peak width of the superlattice intensity above $T_1$ is broader than the instrumental resolution, meaning that the superstructure is short-ranged. The intensity above $T_1$ is therefore given by the fluctuation. The intensity is proportional to the square of the Pd displacement, which must be a good indicator of the Yb valence. Below $T_2$=105~K, the intensity is nearly constant, which means that the charge order amplitude in the LT phase is independent of temperature. The incommensurability varies gradually in the MT phase, and jumps to zero at $T_2$. The absence of the higher harmonic peaks at ($\pm n x$ $\pm n x$ 2.5)$_{\rm t}$ (where $n$ is an integer) and the sudden decrease in the total intensity at $T_2$ with heating indicate that the modulation is sinusoidal.

\section{Discussion}
There are two possible interpretations of the intensity distribution in the MT phase: stripe and checkerboard structures. While we have no decisive data to distinguish the two structures, the stripe model is the more plausible of the two. In Fig.~\ref{fig:map} (a), there are some differences in the intensities of the four peaks. The intensity at $(+x,+x,2.5)_{\rm t}$ is equal to that at $(-x,-x,2.5)_{\rm t}$, and the intensity at $(+x,-x,2.5)_{\rm t}$ is equal to that at $(-x,+x,2.5)_{\rm t}$. However, the former pair is apparently stronger than the latter pair. This difference is accounted for by the domain ratio in the stripe model, while the checkerboard model cannot account for this intensity difference. 

Next, we examine the origins of the incommensurate structure based on the stripe model. 
The valence states of the ions can be treated as Ising pseudospins.
It has long been known that an Ising spin with a ferro-coupling for the nearest neighbor ($J_1>0$) and an antiferro-coupling for the second neighbor ($J_2<0$), i.e., the axial next-nearest neighbor Ising (ANNNI) model, results in an intricate phase diagram that involves wide incommensurate regions and commensurate regions separated by the first order phase transitions.\cite{Bak80PRB,Fisher80PRL} 
The ANNNI model has been applied to $4f$ electron magnets including CeSb\cite{Boehm79PRL}, CeBi\cite{Uimin82JPhys} and UPd$_2$Si$_2$\cite{Honma98JPSJ}, to ferroelectrics such as NaNO$_2$\cite{Selke84ZPhysB} and [N(CH$_3$)$_4$]$_2$$M$Cl$_4$\cite{Shimomura86PRB}, and to charge ordering systems such as NaV$_2$O$_5$\cite{Ohwada01PRL} and EuPtP\cite{Inami10PRB}. Among these materials, EuPtP has significant similarities to YbPd, because both are metallic valence fluctuation compounds. 

Let us compare the known properties of the charge ordering in YbPd with those expected from the ANNNI model. First, the characteristic modulation vector varies
 from (0.07, 0.07, 0) at $T_1$ to (0.06, 0.06, 0) at $T_2$. According to ref.~\onlinecite{Bak80PRB}, the wavelength is stabilized in a narrow temperature range with $-J_2/J_1 \sim 0.27$. When the temperature decreases, the calculation then predicts a first order phase transition to a ferroic arrangement within the $c$-plane, which was observed in our experiments at $T_2$. The ratio $-J_2/J_1$ is controlled by the application of pressure.\cite{Shimomura86PRB,Ohwada01PRL} As shown in the inset of Fig.~\ref{fig:006}(b), the LT phase vanishes around 2~GPa, whereas the MT phase remains above that pressure level. When we simply map the ANNNI phase diagram to the YbPd temperature-pressure phase diagram, then the charge order characterized by the wavevector ($\frac14 \frac14 \frac12$), which corresponds to $2\sqrt{2} a$ in-plane periodicity, is expected for the ground state under pressures of more than 2~GPa. It is also expected that a large number of phase transitions will occur in the MT phase. We can test the applicability of the ANNNI model to YbPd by performing further diffraction measurements in high pressure environments.

Another common origin of the long wavelength modulation is Fermi surface nesting. However, the nesting vector can vary only moderately. The cell volume change at $T_2$ is almost the same as that within the MT phase, which means that the change in the Fermi wavevector at $T_2$ should be at a similar level to the change within the MT phase. As shown in the temperature dependence of the incommensurability $x$ (Fig.~\ref{fig:map}(c)), the jump in $x$ at $T_2$ is far greater than its shift in the MT phase. Therefore, the long wavelength modulation is unlikely to be caused by nesting of the Fermi surface.

Finally, we discuss the effects of the charge ordering on the magnetic and conductive properties of the material. 
 The application of pressure makes the LT phase unstable, and the ground state charge ordering structure is changed from an $x=0$ structure to an $x\neq 0$ structure (the MT structure) above 2~GPa. In accordance with this transition, $T_{\rm N}$ vanishes.\cite{Miyake} Therefore, the magnetic ordering at $T_{\rm N}$ requires a charge ordered structure in the LT phase. The charge ordering characterized by $x=0$ makes Yb$^{3+}$ and Yb$^{2.6+}$ sublattices, which can have different Kondo temperatures. The Yb$^{3+}$ sublattice can form a magnetic order because it is not in the mixed valence state. In the $x \neq 0$ structure in which the sinusoidal charge modulation develops, the Yb$^{3+}$ content is small, and therefore the magnetic ordering is suppressed. 

 In general, charge ordering makes the carrier density decrease, and therefore the resistivity increases, as reported in Yb$_4$As$_3$\cite{Ochiai90JPSJ} and in many charge ordered 3$d$ electron systems.\cite{MItransition} In contrast, the resistivity of YbPd decreases when the charge ordering occurs. This feature originates from the good metallic nature of YbPd. The carrier density is nearly unchanged by the charge ordering because the average valence remains constant, while the randomness of the valence arrangement, which scatters the carriers, is suppressed by the charge ordering.

\section*{Summary}
We have performed a series of x-ray diffraction measurements on the valence fluctuating compound YbPd. The material shows two-fold charge ordering characterized by the wavevector $(00\frac12)_{\rm t}$ below 105~K. Between 105~K and 130~K, the charge ordering structure is modulated, and the characteristic wavevector is $(\pm x, \pm x, \frac12)_{\rm t}$ with $x\sim 0.07$. We propose that this long wavelength structure can be described by using the ANNNI model.

\section*{Acknowledgments}

This work was supported by KAKENHI (Grant No. 23684026), the Japan Securities Scholarship Foundation and the Global COE Program (G10).


\begin{thebibliography}{99}

\bibitem{MItransition} M.~Imada, A.~Fujimori, and Y.~Tokura, Rev. Mod. Phys., {\bf 70} 1039 (1998).


\bibitem{org} H.~Seo, C.~Hotta, and H.~Fukuyama, Chem. Rev. {\bf 104}, 5005
(2004).


\bibitem{Ochiai90JPSJ} A.~Ochiai, T.~Suzuki, and T.~Kasuya, J.Phys. Soc. Jpn. {\bf 59} 4129 (1990). 

\bibitem{Staub05PRB} U.~Staub, M.~Shi, C.~Schulze-Briese, B.D.~Patterson, F.~Fauth, E.~Dooryhee, L.~Soderholm, J.O.~Cross, D.~Mannix, and A.~Ochiai  Phys.Rev. B {\bf 71}, 075115 (2005). 

\bibitem{Inami10PRB} T.~Inami, S.~Michimura, A.~Mitsuda and H.~Wada, Phys. Rev. B {\bf 82} 195133 (2010). 

\bibitem{Pott85PRL} R.~Pott, W.~Boksch, G.~Leson, B.~Politt, H.~Schmidt, A.~Freimuth, K.~Keulerz, J.~Langen, G.~Neumann, F.~Oster, J.~R\"ohler, U.~Walter, P.~Weidner, and D.~Wohlleben, Phys. Rev. Lett. {\bf 54} 481 (1985).

\bibitem{Bonville86PRL} P.~Bonville, J.~Hammann, J.A.~Hodges, P.~Imbert, and G.J.~J\'ehanno, Phys. Rev. Lett. {\bf 57} 2733 (1986).

\bibitem{Iandelli73RevChimMiner} A.~Iandelli and A.~Palenzona, Rev. Chim. Miner. {\bf 10} 303 (1973).

\bibitem{Miyake12JPhysConf} A.~Miyake, T.~Kagayama, K.~Shimizu, and T.~Ebihara, J. Phys.: Conf. Ser. {\bf 391} 012045 (2012). 

\bibitem{Nakanishi11ChineseJPhys} Y.~Nakanishi, T.~Kamiyama, K.~Ito, M.~Nakamura, M.~Sugishima, A.~Mitsuda, H.~Wada, and M. Yoshizawa, Chinese J. Phys. {\bf 49} 462 (2011). 

\bibitem{Canfield92PhilMag} P.C.~Canfield and Z.~Fisk, Phyl. Mag. B {\bf 65} 1117 (1992).

\bibitem{Qd} The scattering amplitude from a crystal with a structural modulation is $F(\vec Q)= \sum_n f_n \exp (i \vec Q \cdot (\vec R_n+\vec \delta _n))\simeq  \sum_n f_n \exp (i \vec Q \cdot \vec R_n)(1+ i \vec Q \cdot \vec \delta _n))$, where $\vec R_n$ and $\vec \delta_n$ denote the average atomic position and atomic displacement of the $n$-th atom. The amplitude proportional to $\vec Q \cdot \vec \delta _n$ corresponds to the superlattice reflection, while the rest of the amplitude corresponds to the Bragg reflection. The superlattice intensity is thus proportional to $|\vec Q \cdot \vec \delta|^2$.

\bibitem{Dumesnil98PRB} K.~Dumesnil, A.~Stunault, Ph.~Mangin, C.~Vettier, D.~Wermeille, N.~Bernhoeft, S.~Langridge, C.~Dufour, and G.~Marchal, Phys. Rev. B {\bf 58} (1998) 3172.

\bibitem{Sasaki_table} S.~Sasaki, KEK Rep. {\bf 88-14}, 1 (1989).

\bibitem{Bak80PRB} P.~Bak and J. von Boehm, Phys. Rev. B {\bf 21} 5297 (1980).

\bibitem{Fisher80PRL} M.E.~Fisher and W.~Selke, Phys. Rev. Lett. {\bf 44} 1502 (1980).

\bibitem{Boehm79PRL} J. von Boehm and P.~Bak, Phys. Rev. Lett {\bf 42} 122 (1979).

\bibitem{Uimin82JPhys} G.~Uimin, J. Physique Lett. {\bf 43}, 665 (1982). 

\bibitem{Honma98JPSJ} T.~Honma, H.~Amitsuka, S.~Yasunami, K.~Tenya, T.~Sakakibara, H.~Mitamura, T.~Goto, G.~Kido, S.~Kawarazaki, Y.~Miyako, K.~Sugiyama and M.~Date, J. Phys. Soc. Jpn. {\bf 67} 1017 (1998).

\bibitem{Selke84ZPhysB} W.~Selke and P.M.~Duxbury, Z. Phys. B {\bf 57} 49 (1984).

\bibitem{Shimomura86PRB} S.~Shimomura, N.~Hamaya  and Y.~Fujii, Phys. Rev. B {\bf 53}, 8975 (1996).

\bibitem{Ohwada01PRL} K.~Ohwada, Y.~Fujii, N.~Takesue, M.~Isobe, Y.~Ueda, H.~Nakao, Y.~Wakabayashi, Y.~Murakami, K.~Ito, Y.~Amemiya, H.~Fujihisa, K.~Aoki, T.~Shobu, Y.~Noda, and N.~Ikeda, Phys. Rev. Lett. {\bf 87}, 086402 (2001).

\bibitem{Miyake} A.~Miyake, K.~Kasano, T.~Kagayama, K.~Shimizu, R.~Takahashi, Y.~Wakabayashi, T.~Kimura and T.~Ebihara, in preparation.

\end{thebibliography}
\end{document}